\documentclass[preprint,proceedings]{rmaa}
\usepackage{rmaacite}
\usepackage{ringhead}

\newcommand{\hi}{{H$\,$\footnotesize I}}
\newcommand{\bb}{$B$}
\newcommand{\wc}{$W_\mathrm{20}^\mathrm{c}$}
\newcommand{\vel}{$v_\mathrm{sys}$}
\newcommand{\sm}{$\sim\,$}
\newcommand{\kms}{\,km\,s$^{-1}$}
\newcommand{\df}{\mbox{DEF}}
\newcommand{\dfbis}{<DEF>}
\newcommand{\fdf}{$F_\mathrm{DEF}$}
\newcommand{\fdef}{$F_\mathrm{DEF}^\mathrm{c}$}
\newcommand{\subhi}{_\mathrm{H\mbox{\tiny I}}}
\newcommand{\mhi}{M\subhi}
\newcommand{\fhi}{F\subhi}
\newcommand{\shi}{\overline{\Sigma}\subhi}
\newcommand{\degr}{^{\circ}}

\newcommand{\aas}{Astronomy \& Astrophysics Suppl.\ Ser.}
\newcommand{\sci}{Science}     
\newcommand{\D}{\discretionary{}{}{}}

\SetYear{2002}
\SetConfTitle{Ringberg Workshop on the Virgo Cluster}

\title{The \hi\ Content of the Virgo Cluster Galaxies}

\author{J.~M.\ Solanes\altaffilmark{1}}
\shortauthor{Solanes} \shorttitle{\hi\ Content of Virgo}

\altaffiltext{1}{Departament d'Enginyeria
Inform\`atica i Ma\-te\-m\`a\-ti\-ques, Universitat Rovira i 
Virgili. Av.\ Pa\"\i sos Catalans, 26; 43007~Ta\-rra\-go\-na, Spain. (jsolanes\D{}@etse.\D{}urv.\D{}es).}

\suppressfulladdresses

\addkeyword{cosmology}
\addkeyword{galaxies: clusters: Virgo}
\addkeyword{galaxies: evolution}
\addkeyword{galaxies: ISM}
\addkeyword{galaxies: spiral}

\begin{document}
\maketitle 

\boldabstract{Because of its proximity and richness on late-type
galaxies, the Virgo cluster has been the subject of numerous studies
exploring the effect of the intracluster environment on galaxy
evolution. First discovered nearly three decades ago by \scite{DL73},
the neutral hydrogen (\hi) deficiency of the Virgo galaxies is now a
well-established observational phenomenon also observed on many other
rich clusters. A number of studies on the neutral hydrogen 21-cm line
of Virgo cluster galaxies ---from the lenticular and largest spiral
types through the faintest dwarf irregulars--- show that a significant
number of these objects have lost a substantial fraction of their
atomic hydrogen and are \hi-deficient when compared with galaxies of
the same optical properties in less dense environments.}

\vspace*{+2mm}
\section{Introduction}

Nowadays, the wealth of \hi\ data gathered for the Virgo cluster region
makes it possible to obtain a reliable description of the pattern of
neutral gas deficiency on supercluster scales. I will present here
a recent evaluation of the large-scale radial run of the \hi\
deficiency on the Virgo~I cluster (VIC) region traced by the giant
spiral population, as well as of the distribution of this property in
the two- and three-dimensional space, obtained from the combination of
21-cm data with Tully-Fisher (TF) distance measurements. I will also
attempt to provide suggestive evidence that objects with high \hi\
deficiencies are not exclusively confined to the Virgo cluster proper,
but can be also observed both in a background galaxy group at \sm
25--30 Mpc from us (possible related to the classical W' cloud) and in
various galaxies lying in the frontside of the cluster at line-of-sight
(LOS) radial distances less than 15 Mpc.

A dynamical model for the collapse and rebound of spherical shells
under the point mass and radial flow approximations will be used to
demonstrate that it is not unfeasible that some galaxies far from the
cluster, including those in the gas-deficient group well to its
background, went through the cluster core a few Gyr ago. The
implications would be: (1) that a substantial fraction of the
\hi-deficient spirals in the VIC region might have been deprived of
their neutral hydrogen by interactions with the hot intracluster
medium; and (2) that objects spending a long time outside the cluster
cores might keep the gas deficient status without significantly
altering their morphology.

\vspace*{+2mm}
\section{Observational Data}\label{data}
\subsection{Sample Selection and Homogenization}

Our data set \scite{Sol02} is based on the complete spiral sample used
by \scite{YFO97} to study the structure of the Virgo cluster from
\bb-band TF distances. This sample has been supplemented by data from
seven other TF studies of the Virgo cluster. We have selected from the
original catalogs only galaxies with heliocentric radial velocities
below the well-defined gap near 3000 \kms\ that neatly isolates the
Virgo region in redshift space \cite{BPT93}. In addition, we have
focused on galaxies located in the region bounded by $\rm 12h\le
R.A.\le 13h$ and $\rm 0\degr\le Decl.\le +25\degr$ (equatorial
coordinates are referred to the B1950.0 equinox), which encompasses the
\emph{Virgo Cluster Catalog} (VCC) survey boundary and is centered on
the classical VIC \cite{dVau61}. The selection procedure implies that
all galaxies included in our catalog are expected to have peculiar
motions influenced by the central mass concentration of the cluster.

The initial selection of Virgo galaxies includes a total of 198 objects
representing virtually all spiral galaxies used to date in the
application of the TF relation to study the VIC region. Since TF
distances from different data sets are not always consistent, the
average of the measurements available for each object does not
necessarily provide the best estimate of the galaxy distances. Thus,
the all catalogs have been reduced to a homogeneous system by
eliminating systematic differences among the different sources by means
of a recursive procedure (see Solanes et al.\ 2002, for details). A
Table, available through the Virgo web page, lists the individual and
homogenized distance moduli for the 198 galaxies initially selected.

\subsection{The 21-cm sample}\label{21cm}

Since irregular and bulge-dominated galaxies may give unreliable TF
distances and/or \hi\ content measurements, the morphologies of the
galaxies have been re-examined after the homogenization of the distance
moduli in order to pick up only those with Hubble types ranging from
$T=1$ (Sa) to $T=9$ (Sm), as given in the \emph{Third Reference
Catalogue of Bright Galaxies} (RC3).

The principal source for the \hi\ line fluxes is the \emph{Arecibo
General Catalog} (AGC), maintained by Riccardo Giovanelli and Martha
P.\ Haynes at Cornell University, which contains an extensive
compilation of 21-cm-line measurements collected from a large number of
sources. \hi\ flux measures for a few of objects were also taken from
\emph{A General Catalog of \hi\ Observations of Galaxies} \cite{HR89}
and from the \emph{Lyon-Meudon Extragalactic Database} (LEDA). All the
observational values have been corrected for the effects of random
pointing errors, source extent, and internal \hi\ absorption following
\scite{HG84}, except the (nine) non-AGC fluxes and the only
non-detection, V0522, which have been corrected only for internal \hi\
self-absorption. The AGC is also adopted as the source of other
observational parameters required by our study, such as the equatorial
coordinates of the galaxies, their visual optical diameters, which are
involved in the determination of the \hi\ deficiency, and their
heliocentric radial velocities, which are transformed to systemic
velocities, \vel, by referring them to the kinematic frame of the Local
Group, taken equal to 308 \kms\ towards $(l,b) = (105\degr,-7\degr)$
\cite{YTS77}.

\begin{figure}[!t]
\vspace*{-2mm}
\includegraphics[bb=34 14 550 582,width=\columnwidth]{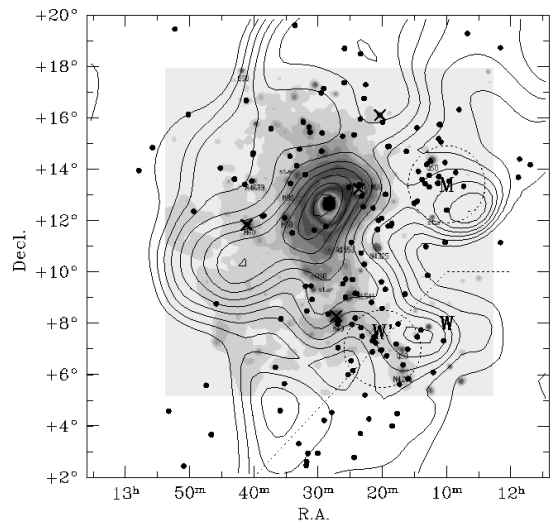}
\vspace*{-10mm} 
\caption{Distribution in celestial coordinates of the 161 members of
the 21-cm sample. The contour map of the \emph{distance-independent}
\hi\ deficiency parameter is reproduced from \scite{Sol01}. A
grey-scaled version of the X-ray image of the central cluster region in
the ROSAT all-sky-survey in the hard (0.4--2.4~keV) energy band is
overlaid also on the figure. The background W, W$^\prime$, and M
subgroups from \scite{BPT93} are delineated by dotted lines. The sky
positions of five dominant galaxies are marked by crosses (top to
bottom: M100, M86, M87, M60, and M49). The projected location of M87
coincides with the peak of the \hi\ deficiency and of the X-ray
emission.}\label{virgohiam}
\end{figure}

On the other hand, for the \hi-line width we have adopted the
inclination-corrected values of the line width at 20\% level of the
line-profile peak, \wc, listed in \scite{YFO97}. For most of the
galaxies that concern us here, these authors provide a set of
observations standardized into line widths measured at the Arecibo
circular feed following a similar process to that carried out in
\scite{Sol02} with the distance moduli. For galaxies not included in
the Yasuda et al.'s sample, we use the values of $W_\mathrm{20}$ and
inclination quoted in LEDA, except for the galaxy V1043, not listed in
either of these two catalogs, for which we adopt the corresponding
measurements by \scite{MAH80}. Furthermore, we have excluded spiral
galaxies with \wc$\;\le 100\,$\kms\ to reduce the error induced from
turbulent disk motion. We do not find it necessary, however, to impose
an inclination cut.

After all these selections, we end up with a sample of 161 spiral
galaxies with reliable \hi\ content and distance data, hereafter called
the ``21-cm sample'', useful to assess the spatial distribution of the
neutral gas deficiency in the VIC region. The scatter of the best
fitting TF template for the 161 galaxies is 0.41~mag. Hence, the
uncertainty in the distance modulus of the individual galaxies in this
data set is comparable to the scatter of the most accurate TF template
relations currently available.

A second table with the most relevant galaxy parameters is also
available through the Virgo web page. The sky distribution of the
members of the 21-cm sample is presented in Figure~\ref{virgohiam}.

\vspace*{+2mm}
\section{The Diagnosis of \hi\ Deficiency}\label{hidef} 

\hi\ deficiency is often quantified by the parameter $\dfbis$ defined as
\begin{equation}\label{def2}
\dfbis\;=\;\langle\log\mhi(D_\mathrm{opt},T)\rangle-\log\mhi\;,
\end{equation} 
\cite{HG84,SGH96}, where $\mhi$ is the \hi\ mass of the galaxy in solar
units, and the angular brackets on the right of the equal sign indicate
the expected value of this quantity inferred from a sample of field
galaxies of the same \emph{optical} linear diameter $D_\mathrm{opt}$
and morphological type $T$.

The neutral hydrogen mass, in turn, is calculated from the expression
\begin{equation}  
\mhi=2.36\times10^5d^2\fhi\;,
\end{equation}
where $d$ is the observed LOS distance of the object in
Mpc and $\fhi$ represents the corrected \hi\ flux density integrated
over the profile width in units of Jy\kms. The most recent
determinations of the expectation values for the \hi\ mass as a
function of the size and morphology of the galaxies are given in
\scite{SGH96} in the form of linear regressions that imply power
law relationships of the type $\mhi\propto D_\mathrm{opt}^n$, with the
values of $n$ oscillating between about 1.7 for Sc's and 1.2 for
earlier spiral types.

It is also possible to use a calibrator for the neutral gas deficiency
not tied to the distance to the galaxies. Given that the
$\mhi-D_\mathrm{opt}$ relationships do not deviate substantially from a
constant \hi\ surface density, especially for the latest spiral types,
it is reasonable to adopt the distance-independent approximation to
equation~(\ref{def2}) based on the difference of the logarithms of the
expected and observed values of this latter quantity
\begin{equation}\label{def1} 
\df=\langle\log\shi (T)\rangle-\log\shi\;,
\end{equation}
where $\shi$ is the mean \emph{hybrid} \hi\ surface density, which can
be calculated directly from the ratio of the observables $\fhi$ and the
apparent optical diameter of the galaxy, $a_\mathrm{opt}^2$, given in
arcmin \cite{SGH96}. The adopted values for
$\langle\log\shi (T)\rangle$ are: 0.24 units for Sa, Sab; 0.38 for Sb;
0.40 for Sbc; 0.34 for Sc; and 0.42 for later spiral types. 

\vspace*{+2mm}
\section{The Distribution of \hi\ Deficiency}\label{hiversusd}

Numerous studies \cite{GH85,HG86,Mag88,Cay94,Bra00} reveal that
gas-poor galaxies tend to be more abundant in the centers of rich
galaxy clusters than in their periphery. Virgo is less rich and younger
than the classical Abell clusters and is characterized by a lower X-ray
luminosity and larger spiral fraction than Coma-like clusters. Probably
as a result, although it does contains a substantial fraction of
\hi-deficient galaxies \cite{HG86,YFO97,Sol01}, the degree of \hi\
deficiency is not observed to increase towards the center as
dramatically as in other rich clusters. However, because of its
proximity, even strongly gas-poor galaxies remain detected, allowing
precise determination of higher degrees of the \hi-deficiency, whereas
in more distant clusters only lower limits to this parameter can be
derived.

The contour map of \hi\ deficiency shown in Figure~\ref{virgohiam}
illustrates that the maximum of the gas deficiency distribution
coincides with the position of the central cD galaxy, M87, where the
projected galaxy and intracluster gas densities are also the
highest. But this map also reveals other zones of significant
deficiency at sky positions dominated by background subclumps which lie
at about twice the distance of the cluster core \cite{TS84,BTS87,Gav99}
and are believed to be falling into it for the first time.

\subsection{Radial Pattern}

\begin{figure}[!t]
\vspace*{-6mm}
\includegraphics[width=\columnwidth]{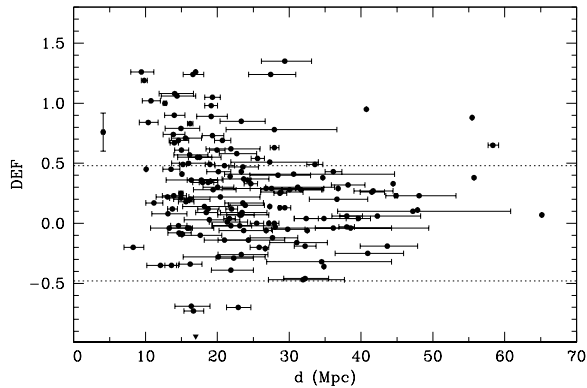} 
\vspace*{-15mm}
\caption{Individual
values of $\df$ for the 161 members of the 21-cm sample as a function
of the LOS distance. Dotted lines show 2 times the standard deviation
shown by the values of this parameter in field galaxies. Horizontal
error bars represent the $1\sigma$ uncertainties of the distances
quoted in the literature with respect to the calculated mean
values. The filled triangle marks the distance to M87 quoted in
LEDA. The vertical error bar in the point closest to us shows an
estimate of the typical uncertainty of the individual values of $\df$
expected from random errors in the determination of the observables
$a_\mathrm{opt}^2$, \fdef, and $T$, that enter in the calculation of
this parameter.}\label{hivsd}
\end{figure}

\begin{figure}[!t]
\vspace*{-6mm}
\includegraphics[width=\columnwidth]{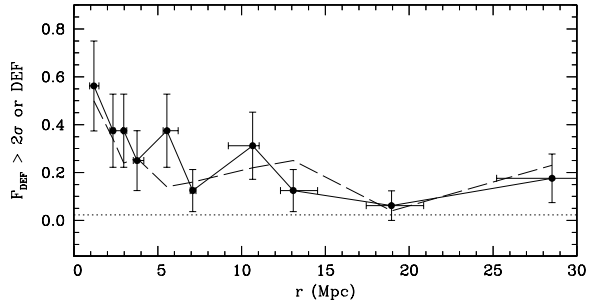}
\vspace*{-15mm} 
\caption{Run of the fraction of spirals with $\df>2\sigma$ with
three-dimensional radial distance from the center of the Virgo
cluster. Vertical error bars correspond to $1\sigma$ confidence Poisson
intervals. The abscissas show medians and interquartile ranges of the
bins in distance determined from 16 galaxies, with the remainder one
added to the last bin. The horizontal dotted line is the expectation
value of \fdf\ for field spirals if $\df$ follows a gaussian
distribution. The long-dashed curve illustrates the radial run of the
medians of the binned number distributions in the measured
$\df$.}\label{hivsd3}
\end{figure}

The first lines of evidence that there are a considerable number of
galaxies with strong \hi\ depletions on the outskirts of the Virgo
cluster are presented by means of Figures~\ref{hivsd} and
\ref{hivsd3}. Figure~\ref{hivsd} shows the values of $\df$ for the 161
members of the 21-cm sample as a function of LOS distance. This diagram
illustrates that most of the galaxies with substantial deficiencies in
the VIC region are localized in a broad range of projected distances,
which stretches from about 10 to 30 Mpc along the LOS. A few more gas
deficient objects lie beyond 40 Mpc.

By transforming the sky positions of the galaxies and their LOS
distances to rectangular coordinates, we can also inspect the behavior
of the \hi\ deficiency as a function of the \emph{three-dimensional}
radial distance, $r$, from the center of Virgo. We adopt the standard
identification of the cluster center at the position of M87, given by
the sky coordinates $(12^{\rm h}28\fm3,12\degr 40\arcmin)$ and a
distance modulus of 31.11 mag quoted in LEDA which translates to a LOS
distance of 16.7 Mpc. The results are shown in Figure~\ref{hivsd3},
where we use two different representations to calculate the radial run
of the \hi\ deficiency: one based on the parameter \fdf, which measures
the relative populations of deficient and normal spirals, and the other
based directly on the averaged values of \hi\ deficiency. In both cases
the data have been binned into annuli containing 16 galaxies per ring,
with the final bin having 17, in order to increase the statistical
weight of the scarcer low- and high-distance objects.

For small values of the 3D clustercentric distance, $r\lesssim 4$ Mpc,
the radial behavior of the gas deficiency is consistent with the
pattern exhibited by the composite sample of 11 \hi-deficient clusters
investigated in \scite{Sol01}: it decreases almost monotonically with
increasing distance from M87, implying that peak of the \hi\ deficiency
distribution coincides with the cluster center. But at greater
distances this tendency is broken by a series of secondary maxima
---more conspicuous in the radial run of \fdf\ due to its higher
sensitivity to localized enhancements---, showing that localized
regions of gas deficiency in the VIC region can also be found well
beyond the typical clustercentric distance of \sm3$h^{-1}$ Mpc where
where this property approaches normalcy in other \hi-deficient clusters
and where the hot X-ray emitting intracluster medium is
concentrated. Be aware, however, of the fact that this same sort of
careful analysis of \hi\ deficiency at large clustercentric distances
has not been performed on other clusters. So, it is not unfeasible that
the differences in the radial pattern can be explained simply by the
bias that arises from Virgo's proximity which leads to (a) much larger
number of 21-cm observations, (b) more stringent values of $\df$, and
(c) more accurate TF distance estimates.

\begin{figure*}[!t]
\includegraphics[width=\textwidth]{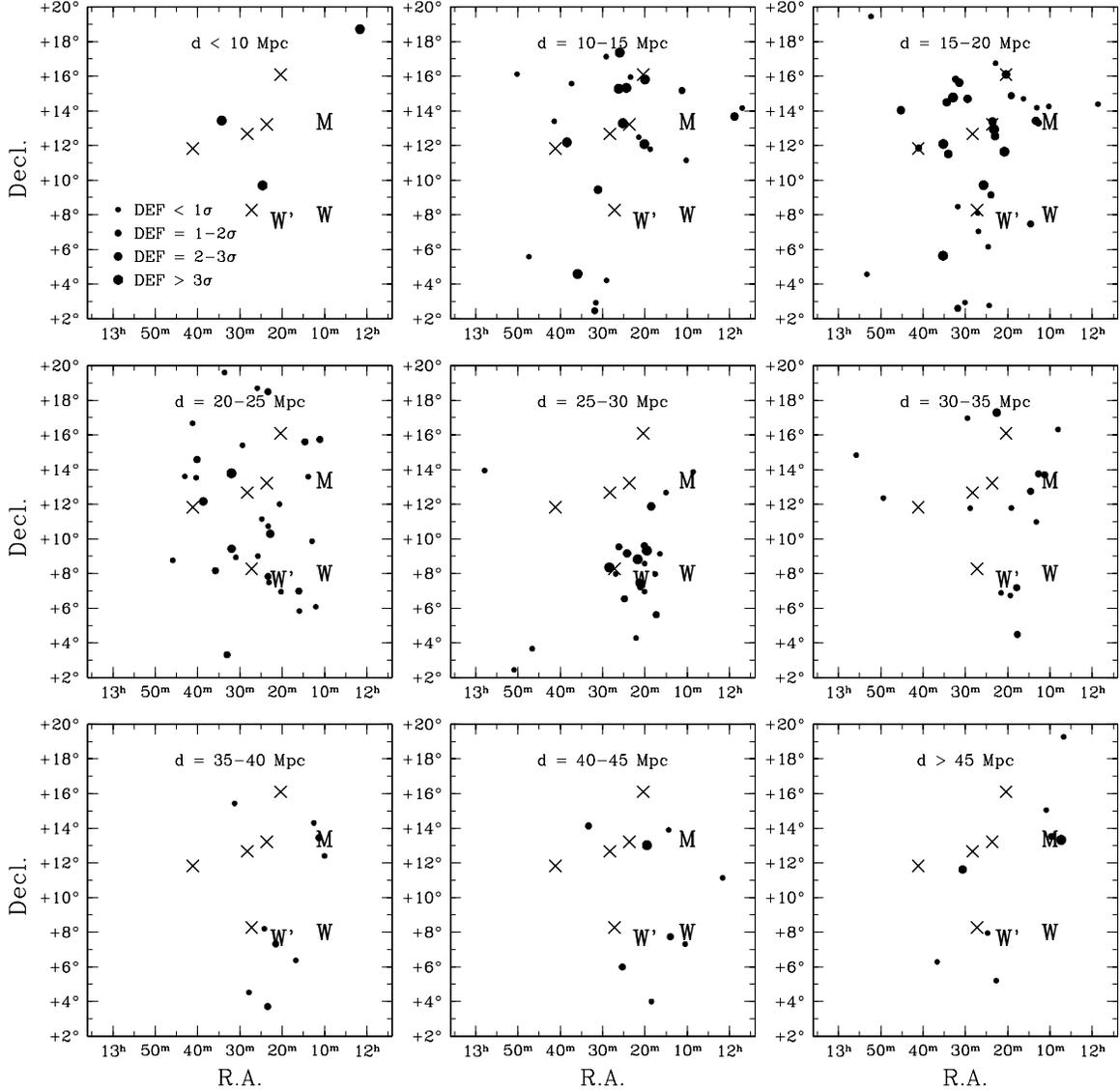} 
\vspace*{-13mm}
\caption{Sky distribution of the Virgo spirals for specific ranges of
the LOS distance. The size of the symbols correlates with the \hi\
deficiency of the galaxies measured in units of the mean standard
deviation for field objects ($=0.24$). Crosses and uppercase letters
have the same meaning as in Fig.~\ref{virgohiam}.}\label{tomoplot}
\end{figure*}

The central peak in the radial pattern of the \hi\ deficiency is
essentially the result of the accumulation of highly deficient galaxies
in the interval of LOS distances ranging from 16--17 Mpc up to 21--22
Mpc. This range coincides with the distribution of the bright
ellipticals associated with the cluster core \cite{NT00}. The second
local maxima visible in the radial run of \fdf\ is produced by galaxies
with LOS distances $\lesssim 15$ Mpc, while the peak most distant from
the cluster core obeys to the grouping of several objects with extreme
deficiencies at LOS distances between about 25 and 30 Mpc from us.

Previous studies by \scite{FOY93,YFO97,FTS98}, among others, have shown
that the Virgo spiral distribution is strongly elongated along the
LOS. The impressions obtained above from the distribution of \hi\
deficiency, although crude, provide further evidence for the large
depth in LOS distance of the Virgo spirals, which we now see that is
also reflected in the gaseous deficiency. Hence, in contrast to what it
is commonly assumed, \emph{not all the \hi-poor objects in the VIC
region reside in the neighborhood of the cluster core}.

\subsection{Three-dimensional Distribution}
\label{structure3d}

Traditionally the determination of the structure of the Virgo cluster
region has relied on the morphological characteristics of its
associated galaxies serving as distance indicators, supplemented by
velocity information \cite{dVau61,VCC,BPT93}. In VCC, for instance, the 
galaxy morphology and membership are strongly coupled, to the point
that, when the membership status is changed (because of the measured
radial velocity) the Hubble type is often redefined. Given the
well-defined kinematical gap behind the Virgo galaxy concentration at
\sm3000 \kms, this technique isolates the cluster region from the (far)
background, but can only provide a fuzzy view of its complex
morphology, especially in regions where several candidate substructures
overlap in projected space and velocity (notably around the western
cluster side, defining what it is known as the ``spoiled'' area).

\begin{figure}[!t]
\vspace*{+8mm}
\includegraphics[bb=130 0 460 770,width=\columnwidth]{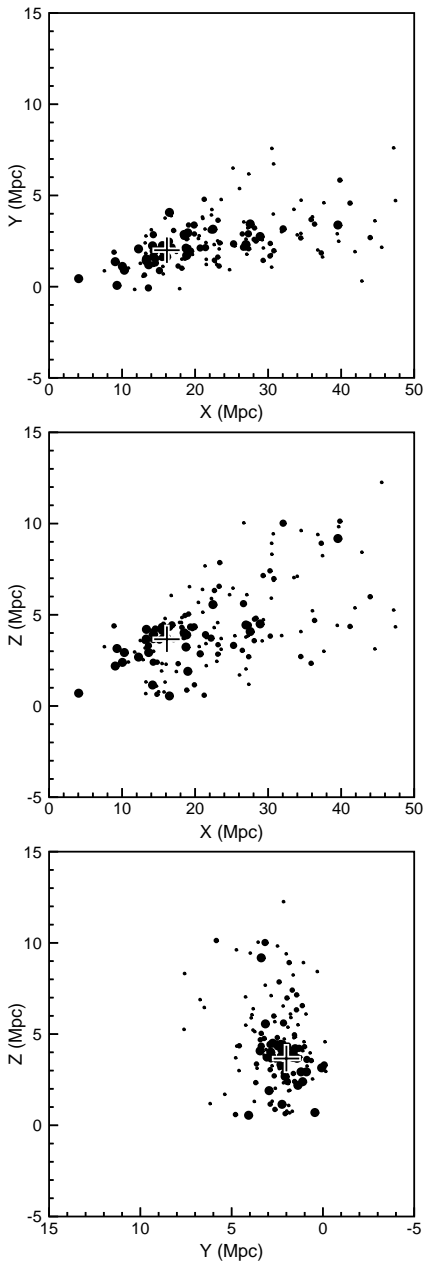}
\vspace*{-20mm}
\caption{Distribution of the VIC spirals in the three main planes of
the rectangular equatorial coordinate system. As in
Figure~\ref{tomoplot}, the symbol size indicates the relative degree of
\hi\ deficiency. The large cross in each panel marks the position of
M87.}\label{cartesian}
\end{figure}

Some progress toward a more precise determination of the internal
structure of the Virgo cluster region and improved membership
assignments is now beginning to emerge from the incorporation of
spatial information based on distance measurement methods capable of
determining individual galaxy distances to a precision comparable to
the inter-group separations. Recent studies relying on TF, SBF, or
fundamental plane distance measurement techniques
\cite{YFO97,Gav99,NT00} have produced quite an elaborate set of
substructures and opened a debate on the original group membership
assignments of numerous galaxies ---a detailed historical account of
the studies on the structure of the Virgo cluster can be found, for
instance, in \scite{Fou01}. In essence, however, they have confirmed
the robustness of the original subdivision inferred from imaging and
recessional velocity data that splits the VIC region essentially in two
major central subclusters and three peripheral groups (cf.\
Fig.~\ref{virgohiam}). The largest galaxy concentration dominates the
northern part of the Virgo region and coincides with the brightest
giant elliptical, M87, which also appears to be the center of the X-ray
emission \cite{Boh94}. This main subunit, which will be referred to
here as the M87 subcluster, is supposed to trace the cluster core,
which might not be virialized. In fact, \scite{BPT93,Boh94,SBB99},
found evidence of dynamical disturbances that could be explained by the
ongoing merging of two galaxy systems: one associated with M87 itself
and the other with the pair of elliptical galaxies M86/M84. Another
giant elliptical, M49, marks the center of the other major Virgo galaxy
concentration, hereafter the M49 subcluster, located southwards from
the M87 subcluster. The M49 subcluster appears to be connected towards
the southwest with the W$^\prime$ and W background clouds
\cite{dVau61}, forming a continuous chain that extends up to roughly
twice the distance of the M87 subcluster (interestingly enough, a
tenuous bridge of X-ray luminous gas can be seen in
Fig.~\ref{virgohiam} connecting the M49 subcluster with the
W$^\prime$/W cloud region). Finally, in the northwest and at about the
distance of the W cloud, there is another well-defined background cloud
named M \cite{FSF84}.

\subsubsection{Spherical Coordinates}

The tomographic presentation of the galaxy distribution shown in
Figure~\ref{tomoplot} clearly demonstrates that the center of gravity of
the \hi\ deficiency distribution moves from north to south as the
distance increases, consistently following the structure of the Virgo
cluster described above. The major concentration of \hi-deficient
spirals is seen in the distance range of 15--20 Mpc encircling the
position of M87. Numerous gas-deficient objects are detected also in
the panels corresponding to the distance ranges of 10--15 Mpc and
25--30 Mpc. In the latter, these galaxies are essentially concentrated
between the southern edge of the M49 subcluster and the W$^\prime$/W
cloud region, while in the former they tend to be located to the north
of M87. Some of the gas poor galaxies in the near distance slice could
be former companions of M86 ejected at high speeds to relatively high
clustercentric distances because of the falling of this subclump into
the cluster \cite{Sol01,Vol01}. The intermediate range of $20<d<25$ Mpc
is composed mainly of galaxies with moderate neutral gas deficiencies
spreaded more or less uniformly over all the sky. Although the
uncertainties in the distance estimates do not permit a neat separation
of the different Virgo substructures, it is interesting to note that
the majority of the objects in the \hi-deficient galaxy clustering seen
at 25--30 Mpc also have systemic velocities not dissimilar from those
of the M87 subcluster, in rough agreement with the original definition
of the W$^\prime$ cloud given in VCC (note that the W cloud is
underrepresented in TF data sets). On the other hand, the marginal
indications of a galaxy enhancement in the NW of the
30--35-Mpc-distance slice might correspond to the M cloud, given that
the candidate galaxies exhibit systemic velocities around 2000
\kms. Beyond 35 Mpc, galaxies become progressively scarce, although
with an apparent tendency to reside in the peripheral W and M cloud
regions. This picture is consistent with the claims that the W and M
background clouds of Virgo are twice as far away as its central
subunits, with the W$^\prime$ cloud being somewhat closer
\cite{BTS87,YFO97}.

A final glance at Figure~\ref{tomoplot} also shows that the
gas-deficient enhancement noted in the 2D image around the region of
the M cloud (cf.\ Fig.~\ref{virgohiam}) is indeed the result of the
chance superposition along the LOS of several spirals with substantial
gas deficiency, but located at very different LOS distances and without
any physical connection.

\subsubsection{Rectangular Coordinates}

\begin{figure}[!t]
\includegraphics[width=\columnwidth]{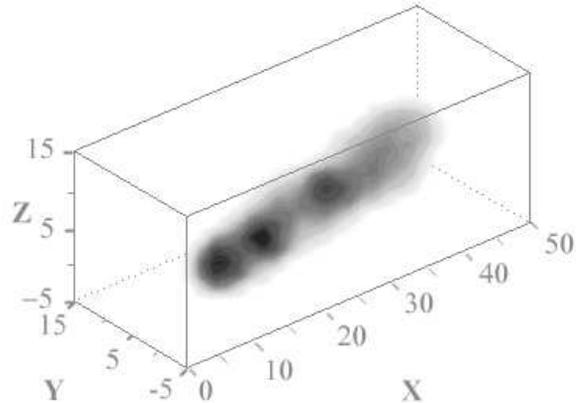}  
\caption{Voxel projection of the 3D distribution of \hi\ deficiency in
the VIC cluster region. The plot is in rectangular equatorial
coordinates. Distances are in Mpc. The xy-plane corresponds to $\rm
Decl.=0\degr$, the x- and y-axis point to $\rm R.A.=12h$ and 18h,
respectively, and the z-axis points to the north. The central
enhancement is associated with the cluster (M87 is right at its
center). We are at the origin of the coordinate system.}\label{voxel}
\end{figure}

A complementary characterization of the spatial structure of the VIC
region can be inferred from the projected distributions of the spiral
galaxies into the three main planes of the cartesian three-dimensional
space visualized in Figure~\ref{cartesian}. In this plot, the xy-plane
is taken parallel to the equatorial plane ($\rm Decl.=0\degr$), with
the x- and y-axis pointing to $\rm R.A.=12h$ and 18h, respectively, and
the z-axis pointing to the north ---as a result, the yz-plane is nearly
perpendicular to the LOS to M87, i.e., it is roughly a tangent plane to
the celestial sphere. The figure allows one to appreciate the true
aspect of the spiral distribution in the VIC region. The most
remarkable feature is the double filamentary structure that can be
distinguished in the xz-plane, which is roughly perpendicular to the
plane of the Local Supercluster. Notice that the galaxies with the
strongest \hi\ depletions, i.e., those with $\df>2\sigma$, or
equivalently, a factor three reduction in the \hi\ mass, delineate the
upper filament that goes through M87 and embraces the deficient objects
having the most extremal radial distances ---they are indeed much more
densely concentrated along the filament axis than the rest. This upper
branch of the spiral distribution is pretty well aligned with the chain
of bright elliptical galaxies that defines the principal axis of Virgo
\cite{Arp68,WB00}. Emerging from this branch at around 25--30 Mpc in
LOS distance there is a second filament essentially devoid of highly
deficient objects.

A continuous representation in rectangular equatorial coordinates of
the spatial distribution of \hi\ deficiency in the VIC region is shown
in Figure~\ref{voxel}. This image is like a radiography in which the
shade intensity informs on the average \hi\ deficiency of the galaxy
distribution observed under a given viewing angle. Three dark spots
indicating the accumulations of galaxies with a dearth of neutral
hydrogen described above are easily identified aligned along the LOS.
While the frontside enhancement of the \hi\ deficiency is produced by
gas-poor galaxies that appear relatively clustered in 3D space simply
because they are nearby objects, the gas-deficient enhancement in the
background arises from a compact aggregation of galaxies in the
four-dimensional position-radial velocity phase space which is clearly
differentiated from its surroundings. Up to 15 of the galaxies listed
in the 21-cm sample are probable members of this background group. They
all share similar positions in the plane of the sky ($\rm 12h15m\le
R.A.\le 12h30m$\ and\ $\rm +6\degr\le Decl.\le +10\degr$) where
uncertainties are negligible. In addition, 12 objects have LOS
distances between 27 and 30 Mpc, and 8 of those (11 out of the initial
15) have systemic velocities between \sm600--1300 \kms. Certainly, the
lack of resolution in the radial direction prevent us for claiming that
we have identified a true group on a sufficiently safe basis. Yet, the
fact that one third of the candidate galaxies have gas deficiencies
that deviate more than $2\sigma$ from normalcy and that two of them
have \hi\ masses less than 10\% of the expectation values for their
morphology ---characteristics that are both typical of rich cluster
interiors--- reinforces the impression that the compactness of these
objects in the phase space is not fortuitous.

\vspace*{+2mm}
\section{The Radial Velocity Field}\label{predictions}

\begin{figure*}[!t]
\vspace*{-42mm}
\includegraphics[width=\textwidth]{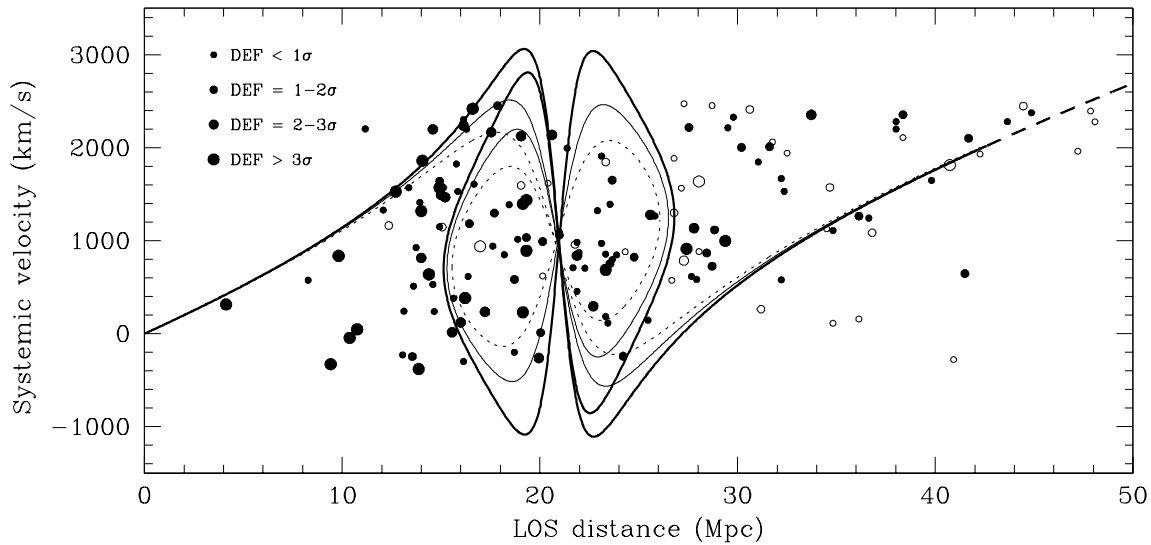}
\vspace*{-47mm} 
\caption[fig2.eps]{Systemic velocity vs.\ LOS distance for spirals
listed in the 21-cm sample. Solid circles identify objects with
reliable distance estimates. The curves demonstrate the predicted
velocities for $\theta=4\degr$ (thick solid), $\theta=6\degr$ (thin
solid), and $\theta=8\degr$ (dotted). The dashed portion is for unbound
shells. Galaxies with uncertain distances (open symbols) have been
excluded from the fit.}\label{hubdiam}
\end{figure*}

The systemic velocity-distance diagram for the VIC region plotted in
Figure~\ref{hubdiam} insinuates the basic expected features: an initial
steeply rising velocity-distance relation at the cluster front, a
central very broad region with the maximum observed velocity
amplitudes, and a final ascending part of the relation, expected to
approach asymptotically the local Hubble law. Ultimately, all the
available good-quality observations should be considered to define the
constraints of any dynamical model of the virgocentric velocity
field. However, given the difficulties inherent to the modeling of the
motions in the innermost cluster region where multiple rebounds are
expected to occur, the acceptable range of models has been restricted
by putting all the weight of the fits in the the two asymptotic
branches of the envelope to the streaming motions with respect to the
Virgo velocity. The theoretical prediction has been derived from the
simple point mass model developed in \scite{San02} for the spherical
collapse of a zero-pressure fluid, which extends the calculations
beyond the time until a singularity first develops until galaxies
recollapse again. This is done by taking the, admittedly crude, point
of view of treating galaxies as test particles moving under the
influence of a central constant potential well, so the effects of shell
crossing on their first orbit following rebound can be ignored.  Two of
the parameters of the model are allowed to vary freely: the
\emph{effective} total mass of the VIC region, $M_{\mathrm{VIC}}$, and
the barycentric distance from our position, whereas the velocity of the
VIC in the Local Group (LG) reference frame has been fixed to 980 \kms\
\cite{Tee92}.

The three solid curves in Figure~\ref{hubdiam} demonstrate the locus of
the optimum model for lines-of-sight corresponding to angular
separations of 4, 6, and 8 degrees. It is compelling that, in spite of
the fact that the minimization relies only on a fraction of the body of
the data, the best solution is able to explain a great deal of the
galaxy motions around the VIC region. The good accordance between the
observations and the prediction of the double-infall model is
reinforced by the remarkable symmetry of the motions with respect to
the local Hubble flow defined by the imaginary straight line going
through the position of the LG and that of the predicted VIC
barycenter. On the other hand, the best estimate
$M_{\mathrm{VIC}}=2.8\times 10^{15}$ M$_\odot$ approaches closely the
values $\lesssim 2\times 10^{15}$ M$_\odot$ inferred from modelings of
the velocity field of the Local Supercluster \cite{TS84,Fou01}. The
acceptable barycentric distances are also well within the very poorly
constrained range ($16-24$ Mpc) of VIC distances reported in the
literature, although the most likely value $R_{\mathrm{VIC}}=21.0$ Mpc
advocates a large-distance scale which, for a cosmological VIC velocity
of 1200 \kms, brings the \emph{local} value of the Hubble constant to
about 60 km/s/Mpc. Furthermore, the best solution leads to a cosmic age
$t_0=13.5$ Gyr, in excellent agreement with the expansion ages of
$13\pm 1$ Gyr derived for the $\Omega_\mathrm{m}=0.3$,
$\Omega_\Lambda=0.7$ CDM model in the $H_0$-Key Project by
\scite{Fre01} and the more precise $t_0=13.6\pm 0.2$ Gyr recently
inferred by \scite{Sie02} from CMB observations.

Interestingly enough, the \hi-deficient background group observed at
$R\sim 28$ Mpc and $V\sim 1000$ \kms can be accommodated the same on
an orbit following rebound as on a first infall trajectory. One would
be therefore compelled to conclude that it is not unfeasible that these
galaxies are not recent arrivals but have already plunged into the VIC
center in the past. The near turnaround position of this entity in the
Hubble diagram and the assumed standard harmonic oscillation movement
of the galaxies around the cluster center, together with a cosmic age
presumably close to 13.5 Gyr, indicate that it might have experienced
its first high-velocity passage\footnote{Galaxies on first infall
achieve much higher pericentric velocities than after relaxation
because their turnaround radius is significantly larger. Hence they can
be subject to a much higher ram pressure.} through the Virgo core about
4.5 Gyr ago.

\vspace*{+2mm}
\section{Concluding Remarks}\label{conclusions}

\begin{itemize}

\item Further progress in the knowledge of the detailed structure of the
Virgo cluster needs a careful revision of TF distances ---at least
until Cepheid distance measurements in Virgo galaxies become more
commonplace. Even after the elimination of systematic differences among
published Virgo catalogs, a few galaxies still exhibit strongly
inconsistent distance measurements: 16 of the 161 members of the 21-cm
sample have $1\sigma$ uncertainties larger than 5 Mpc. The fact that
the quoted error is small does not guarantee that the measurement is
reliable. If two results clearly disagree, any average, weighted or
not, is meaningless, and there is little point in performing it.

\item There is now compelling evidence of the decisive participation of
ram-pressure stripping, which requires high IGM densities and relative
velocities, in the reduced gas abundances of the spirals observed in
the centers of various rich clusters, either from observational
articles \cite{GJ87,DG91,Sol01} ---including the discovery of galaxies
with shrunken \hi\ disks \cite{Cay94,Bra00}--- or from theoretical
studies \cite{SS92,FN99,SAP99,QMB00,Vol01}. The finding that a number
of spirals with substantial \hi\ deficiencies lie at large radial
distances from the Virgo cluster center ---some are very likely members
of a background subclump well behind the cluster core--- may seem from
the outset hard to reconcile with the proposition that this
environmental process is also the cause of their gas deficiencies.
However, the modeling of the velocity field around the VIC region
demonstrates that characteristics such as a large virgocentric distance
or a near turnaround position are not by themselves conclusive
indications of a recent arrival. The substantial \hi\ deficiency of the
background subclump found in \scite{Sol02} may well have originated on
an earlier passage of this entity through the Virgo core. A more
precise knowledge of the velocity field around the VIC appears to be
required to confirm or discount this possibility.

\item Even if the tentative suggestion that the \hi-deficient group on
the backside of the VIC might not be a recent arrival is finally proven
well-founded, it is still necessary finding a sensible explanation for
the apparently long time (\sm 4--5 Gyr) its gas-poor members have
maintained a substantial dearth of gas during without noticeable
consequences on their morphologies: 9 of its 15 probable members are
late-type spirals, whereas the 5 galaxies with the largest gaseous
deficiencies have types Sb or later. Certainly, the details and
chronology of the evolution of galactic properties triggered by the
sweeping of the atomic hydrogen, as well as its repercussions on the
star formation rate, are poorly understood.

\end{itemize}

The unconspicuous positions of the \hi-deficient galaxies observed at
large virgocentric distances ---the gas-deficient group in the
background, for instance, lies halfway between the much larger galaxy
concentrations of the Virgo cluster and the background W cloud--- have
not encouraged systematic 21-cm surveys outside the central VIC
region. I hope that this contribution provides enough grounds for
changing the situation and putting the gas content of the galaxies on
the outskirts of the Virgo cluster under close examination.

\end{document}